\documentclass[final,5p,times]{elsarticle}
\usepackage[colorlinks=true]{hyperref}

\usepackage{amsmath}
\usepackage{amssymb}
\usepackage{mathtools}
\usepackage[utf8]{inputenc}

\newcommand{\id}[1]{\textcolor{blue}{#1}}

\newcommand{\e}{{\mathrm{e}}}

\renewcommand{\d}{\partial}
\renewcommand{\l}{\left(}
\renewcommand{\r}{\right)}
\newcommand{\la}{\langle }
\newcommand{\ra}{\rangle }
\newcommand{\be}{\begin{equation}}
\newcommand{\ee}{\end{equation}}
\newcommand{\ba}{\begin{align}}
\newcommand{\ea}{\end{align}}
\newcommand{\bg}{\begin{gather}}
\newcommand{\eg}{\end{gather}}
\newcommand{\bseq}{\begin{subequations}}
\newcommand{\eseq}{\end{subequations}}

\newcommand{\half}{\frac{1}{2}}

\newcommand{\CLns}{{\tt ${\mathcal C}$osmo${\mathcal L}$attice}}

\begin{document}

\title{Fixing IR tail of gravitational waves from domain walls} 

\author[inr,msu]{Ivan Dankovsky}
\ead{ivan.dankovsky@gmail.com}

\author[inr,mpti]{Dmitry Gorbunov}
\ead{gorby@ms2.inr.ac.ru}

\address[inr]{Institute for Nuclear Research of Russian Academy of Sciences, 117312 Moscow, Russia}
\address[msu]{Faculty of Physics, Lomonosov Moscow State University, 119991 Moscow, Russia} 
\address[mpti]{Moscow Institute of Physics and Technology, 141700 Dolgoprudny, Russia}

\begin{abstract}
Numerical simulations of gravitational waves (GW) production during violent evolution of matter inhomogeneities in the early Universe yield highly wiggle infrared parts of the spectra. Obtained directly from the two-point correlation function, these wiggles are nonphysical, corresponding to the parasitic, double frequency terms in naively averaged squared oscillation amplitudes of GW, and hence must be washed out. 
The deep infrared behavior can be predicted on general grounds, e.g. fixed by causality considerations. However, what matters for real observations, e.g. like that of NANOGrav\,\cite{NANOGrav:2023gor}, and what can be  only inferred from numerical simulations, is the infrared slope near the maximum of the spectrum. It reflects the dynamics responsible for the GW production in its heyday, and hence must be accurately predicted.  We illustrate the problem with numerical simulations of the Domain Wall network performed with the help of code \CLns\,\cite{Figueroa:2021yhd}. 
We suggest a numerical procedure to smooth out these parasitic wiggles, which allows us to recover the true spectrum. Being quite generic, it may be applied to numerical simulations of other hypothetical sources of GW possibly operating in the early Universe. The procedure requires to extend the simulation by a few Hubble times after termination of the GW production. We checked with numerical simulations, that the technically natural long-time extension of simulations, which becomes available via  artificial scaling of different parts in the scalar sector equations provided by PRS prescription\,\cite{Press:1989yh}, gives wrong GW spectra even if the source terms are properly rescaled.      
\end{abstract}
\date{}

\maketitle


{\bf 1.} Gravitational waves (GWs) propagate in the Universe without noticeable attenuation except of overall degrading of the amplitude due to a geometrical factor for point sources and except of redshifting and dilution due to the Universe expansion. This feature makes GW a unique tracer of violent processes (if any) in the early Universe. If observed, they would allow one to look back over the past as early as the inflationary epoch \cite{Grishchuk:1974ny,Starobinsky:1979ty}. 

In the Standard Model of particle physics (SM) all the particles, while ultrarelativistic, remain in equilibrium in the primordial plasma of expanding Universe, see e.g.\,\cite{Rubakov:2017xzr}. Once a specie decouples from the plasma it may keep information of what happened in the Universe before the decoupling. And to transfer the information to present observers it must be stable at the cosmological time scales. Among elementary particles of the SM only neutrinos and photons fulfill these requirements and help us to investigate the dynamics of primordial nucleosynthesis and recombination. 

Within the SM we have no experimental tracers to highlight the history of the Universe before neutrino decoupling, except GW. Unfortunately, within the SM and General Relativity there are no sources in the early Universe capable of producing the GW. Fortunately, the SM (and possibly General Relativity) is not complete, since it can not explain neutrino oscillations, dark matter phenomena, baryon asymmetry of the Universe, horizon problem, etc. And some of the suggested extensions predict GW production e.g., during   cosmological phase transitions, due to collapses of topological defects, at quasi de Sitter stage of the Universe expansion etc, for a review see e.g.\cite{Caprini:2018mtu}. The spectrum of GW contains information on processes responsible for GW production providing details on new physics, including those operating at the high energy scales beyond the reach of present (and foreseen) direct experiments.   

Production of GW typically involves some highly violent process, and analytical estimates yield very limited information on the spectra, like position of the maxima. To obtain the shape of the spectra one must resort to the numerical simulation performed for the same sort of  violent process (e.g. a first order phase transition, evolution of a string or domain wall network, etc) and rescale its model parameters to the values predicted (or expected) in the  particular extension of the SM. Accurate numerical simulations is a necessary step in the analysis of a realistic  signal of GW, as we have at LIGO, Virgo and KARGO GW interferometers\,\cite{KAGRA:2023pio,LIGOScientific:2025slb}, which routinely register signals of mergers of black hole binaries.  

Numerical simulations seem to be non-avoidable when attempting to explain the anomalous correlated delays in pulsar timing signal observed by presently operating  pulsar timing arrays\,\cite{NANOGrav:2023gor,Reardon:2023gzh,Xu:2023wog,EPTA:2023fyk}. The published forms of the GW spectrum are fitted with a simple power-law, which parameters exhibit noticeable scatter. However, new data, and future refined joint analysis, along the lines  of Refs.\,\cite{InternationalPulsarTimingArray:2023mzf,Agazie:2024kdi}, hopefully, will tame this behavior and pin down the parameters describing the inferred  GW spectrum. In response to that, numerical simulations of GW production in the early Universe must improve the predictions of GW spectra produced by particular hypothetical sources in the early Universe.     

In this paper we investigate how to get rid of parasitic wiggles, observed in GW spectra calculated in numerical simulations. We consider the infrared part of the spectrum close to its maximum. First, this part is interesting because of the result of pulsar timing arrays, which seem to observe a growing with frequency spectrum. Second, in numerical calculations we can explore quite a limited range of momenta, and so the range near the maximum of the spectra is of the highest priority. Typically this part of the spectra is most sensitive to the dynamics of switching off the GW source and can be useful in distinguishing different mechanisms possibly operating in the early Universe.

We illustrate the "wiggle-problem" with an example of a cosmic Domain Wall (DW) network evolved with the help of numerical code \CLns \cite{Figueroa:2021yhd,Figueroa:2023xmq}. We consider several ways including the extended time interval of simulations provided by PRS-scaling\,\cite{Press:1989yh}, which turns out to yield wrong GW spectra. Intead, we suggest a simple method to achieve our goal, based on separate averaging of each mode, and obtain smooth spectra out of wiggle numerical results. It is quite generic and may be used to improve the result of numerical simulation of GW production by other sources and in other extensions of the SM.    

{\bf 2.} In the expanding Universe the GW are sourced by the transverse
traceless part of the matter energy-momentum tensor. In the conformal
coordinates the metric reads
\[
ds^2=g_{\mu\nu}dx^\mu dx^\nu=a^2(\tau)\l d\tau^2-\l \delta_{ij}+h_{ij}^{TT}({\bf x},\tau)\r dx^idx^j\r,
\]
and the Fourier modes of the tensor perturbations,
\[
h_{ij}^{TT}({\bf k},\tau)=\int \frac{d^3{\bf x}}{\l
    2\pi\r^3}\,\e^{-i{\bf kx}}h^{TT}_{ij}({\bf x},\tau), 
\]
obey the following equation,
\begin{equation}
  \label{GW-equation}
  h_{ij}''+2\frac{a'}{a}h_{ij}'+{\bf k}^2h_{ij}=16\pi G a^2 \Pi^{TT}_{ij}\,.
\end{equation}
Hereafter $G$ stands for the Newton gravitational constant,
primes refer to the derivatives with respect to conformal
time $\tau$, $a(\tau)$ is the scale factor of the expanding Universe
and the source $\Pi^{TT}_{ij}$ is proportional to the Fourier component of the transverse
traceless part of the entire matter energy-momentum tensor, namely 
$\Pi_{ij}a^2\equiv T_{ij}^{TT}({\bf k},\tau)$.

With natural initial conditions $h_{ij}({\bf k},\tau_i)=0$,
$h_{ij}'({\bf k},\tau_i)=0$ one can find the analytic form of the
solution of eq.\,\eqref{GW-equation} at any stage of the Universe
expansion. Assuming the radiation dominating stage, when $a\propto\tau$,
one obtains
\begin{equation}
  \label{GW-solution}
  h_{ij}({\bf k},\tau)=\frac{16\pi\,G}{k\, a(\tau)}\!\!\int_{\tau_i}^\tau\!\!\!
  d\tau'\,a^3(\tau')\,\sin\l k(\tau-\tau')\r\,\Pi_{ij}\l {\bf k},\tau'\r 
\end{equation}
with $k\equiv |{\bf k}|$. 
In numerical simulations $h_{ij}$ are obtained from the numerical solution
of equation \eqref{GW-equation} on a lattice. 

The energy density of GW reads
\begin{equation}
  \label{GW-energy}
\rho_{GW}({\bf x},\tau)=\frac{1}{32\pi Ga^2}\la h_{ij}'({\bf
  x},\tau)h_{ij}'({\bf x},\tau) \ra,
\end{equation}
where averaging goes over several GW periods (wavelength), see
textbook \cite{Maggiore:2007ulw} for discussions of various subtleties associated with the 
gauge invariance, definition of waves, etc. Once
we are interested in stochastic GW background, we also average over
multiple realizations of the source. Numerically, the latter is
performed by running the code with different initial seeds. For
example, if the source $\Pi_{ij}$ is related to some violent dynamics
of a scalar field, one can start at $\tau=\tau_i$ with a specific ``classical''
configuration augmented with fluctuations described by the equal-time correlation functions
\begin{equation}
 \label{general-fluctuations}
 \la \phi({\bf k})\phi({\bf q})\ra\!=\!
 A(k)\frac{\delta({\bf k}+{\bf q})}{\l 2\pi\r^3},\;\; 
 \la \phi'({\bf k})\phi'({\bf q})\ra\!=\!
 B(k)\frac{\delta({\bf k}+{\bf q})}{(2\pi)^3}. 
\end{equation}
In the homogeneous isotropic Universe the functions $A(k)$ and $B(k)$ depend only on the value of the conformal 3-momentum and are determined by the matter dynamics in the earlier epochs. Among the generic choices are quantum fluctuations, when 
\begin{equation}
 \label{quantum-fluctuations}
 A(k)=\frac{1}{2k}\,\Theta (k-k_{cut})\,,\;\;\;\; 
  B(k)=\frac{k}{2}\,\Theta (k-k_{cut})\,,
\end{equation}
with the upper cut on the 3-momentum $k_{cut}$, and thermal fluctuations   
\begin{equation}
 \label{thermal-fluctuations}
 A(k)=\frac{1}{k}\,\frac{1}{\e^{k/T}-1}\,,\;\;\;\;
 B(k)=\frac{k}{\e^{k/T}-1}\,,
\end{equation}
defined by the temperature $T$ that may or may not coincide with the cosmic plasma temperature. 

In real numerical simulations the range of available 3-momenta is limited by the lattice both from above and from below. Moreover, the infrared mode evolution additionally suffers from an artificial impact of periodical boundary conditions on the lattice. It was demonstrated in the case of DW network \cite{Dankovsky:2025pjg}, that the scalar 
configuration in the scaling regime is universal and does  not depend on the initial conditions, i.e. on the choice of the scalar fluctuations. In the numerical simulations this can be achieved with initial fluctuations, parametrized as follows   
\begin{equation}
 \label{general-IC}
 A(k)=\frac{\e^{-\frac{k_{IR}^2}{k^2}}}{k}\,\frac{\Theta (k-k_{UV})}{\l\e^{k/T}-1\r^\alpha}\,,\;\;\;\;
 B(k)=k\,\e^{-\frac{k_{IR}^2}{k^2}}\frac{\Theta (k-k_{UV})}{\l\e^{k/T}-1\r^\alpha}\,,
\end{equation}
and properly chosen values of the parameters $\alpha$, $T$, $k_{UV}$ and $k_{IR}$.

The different seeds means adopting in each simulation different chance Gaussian variables $|\phi({\bf k})|$ with dispersion distributed according to eq.\,\eqref{general-fluctuations} and with a complex phase of $\phi$ normally distributed in $(0\,2\pi]$. 

In analytical estimates, the stochastic nature generically
implies that we know (some features of) the 2-point correlation
function $\la \Pi_{ij}({\bf k},\tau_1)\Pi_{ij}({\bf q},\tau_2)\ra$, which
sources the GW spectrum. Indeed, inserting \eqref{GW-solution} into
\eqref{GW-energy} one obtains for the GW energy density
\begin{equation}
  \label{Ops}
  \begin{split}
&\rho_{GW}=\frac{8\pi G}{a^4(\tau)}\!\!\int_{\tau_i}^\tau\!\!\!\! d\tau_1 a^3(\tau_1)
  \!\!\int_{\tau_i}^\tau\!\!\!\! d\tau_2 a^3(\tau_2)\!\!\int\!\!  d^3{\bf k}d^3{\bf
  q}
  \e^{i({\bf k}+{\bf q}){\bf x}}
  \\&\times\la \Pi_{ij}({\bf k},\tau_1)\Pi_{ij}({\bf q},\tau_2) \ra
\times \!\Biggl(
\frac{\sin{k(\tau\!-\!\tau_1)}\sin{q(\tau\!-\!\tau_2)}}{kq\tau^2}
\!\\ &
-\frac{\sin{k(\tau\!-\!\tau_1)}\cos{q(\tau-\tau_2)}}
{k\tau}
-\frac{\sin{q(\tau\!-\!\tau_2)}\cos{k(\tau\!-\!\tau_1)}}{q\tau}
\\&
+\cos{k(\tau\!-\!\tau_1)}\cos{q(\tau\!-\!\tau_2)}
\Biggr)\,,
    \end{split}
\end{equation}
where we keep only the averaging over the source. In the homogeneous
and isotropic Universe the 2-point correlation function of a
stochastic source can be expressed as
\[
\la \Pi_{ij}({\bf k},\tau_1)\Pi_{ij}({\bf q},\tau_2)\ra=\delta({\bf
  k}+{\bf q})\,{\cal P}(q,\tau_1,\tau_2)\,,
\]
and the GW energy density is constant in space. 
Then from eq.\,\eqref{Ops} one infers the expression for the GW spectrum 
\begin{equation}
  \label{GW-theory}
\begin{split}
\frac{d\rho_{GW}}{d\log k} & = \frac{128\pi^2G k^3}{a^4(\tau)}
\!\!\int_{\tau_i}^\tau \!\!\!\! d\tau_1 a^3(\tau_1)\!\!\int_{\tau_i}^\tau
\!\!\!\! d\tau_2 a^3(\tau_2) \,{\cal P}(k,\tau_1,\tau_2)\\
&\times
\Biggl(
\frac{\sin{k(\tau\!-\!\tau_1)}\sin{k(\tau\!-\!\tau_2)}}{k^2\tau^2}
\!-\!\frac{\sin{k(2\tau\!-\!\tau_1\!-\!\tau_2)}}{k\tau}
\\&
+\!\cos{k(\tau\!-\!\tau_1)}\cos{k(\tau\!-\!\tau_2)}
\Biggr).
\end{split}  
\end{equation}  

Formula \eqref{GW-theory} makes explicit the problem we
intend to solve: it contains several oscillating terms, and
integration over times $\tau_i\leq\tau_1,\tau_2\leq\tau$ can not smooth the
expected wiggles in the spectrum at least for the infrared part,
$k\tau\lesssim 1$. Knowing the correlation function ${\cal P}(k,\tau_1,\tau_2)$, we
can perform the integration and then drop the oscillating term or
average them by hand. 

However, in numerical simulations the spectrum
is inferred directly from eq.\,\eqref{GW-energy} with 
tensor perturbations calculated on the lattice. 
The 3-dimensional lattice contains $ N^3$ sites in total, labeled by
\begin{equation}
\mathbf{n} = (n_1, n_2, n_3), \;\; \text{with}\;\;\; n_i = 0, 1, \dots, N - 1, \quad i = 1, 2, 3. 
\end{equation}
In numerical simulations below we use the lattice with $N=1024$. 
The reciprocal lattice representing Fourier modes is also periodic and discretized in a 3-dimensional lattice. The Fourier modes live in the sites of the reciprocal lattice, which we label as
\begin{multline*}
\tilde{\mathbf{n}} = (\tilde{n}_1, \tilde{n}_2, \tilde{n}_3), \\ \text{with} \quad 
\tilde{n}_i = -\frac{N}{2} + 1, -\frac{N}{2} + 2, \dots, \frac{N}{2} - 1, \frac{N}{2}, \quad i = 1, 2, 3.
\end{multline*}
The discrete Fourier transform (DFT) is defined as
\begin{equation}
f(\mathbf{n}) = \frac{1}{N^3} \sum_{\tilde{\mathbf{n}}} e^{\frac{2\pi i}{N} \tilde{\mathbf{n}}\cdot\mathbf{n}} f(\tilde{\mathbf{n}})\,, \qquad 
f(\tilde{\mathbf{n}}) = \sum_{\mathbf{n}} e^{-\frac{2\pi i}{N} \tilde{\mathbf{n}}\cdot\mathbf{n}} f(\mathbf{n})\,,
\end{equation}
and we distinguish between a function and its Fourier transform only by their arguments. Finally, note that there is a minimum momentum in the reciprocal lattice,
$ k_{\text{IR}} = \frac{2\pi}{L} $, which defines an infrared cutoff scale for the lattice.
The energy density power spectrum of GWs is then computed with the discrete equivalent of Eq.~\eqref{GW-energy}:
\begin{equation}
\begin{split}
\rho_{\text{GW}}(t) &= \frac{1}{32\pi\,GN^3} \sum_{\mathbf{n}} {h'}_{ij}(\mathbf{n}, \tau) {h'}_{ij}(\mathbf{n}, \tau) 
\\&= \frac{1}{32\pi\,G} \frac{1}{N^6} \sum_{\tilde{\mathbf{n}}} {h'}_{ij}(\tilde{\mathbf{n}}, \tau) h_{ij}^{'*}(\tilde{\mathbf{n}}, \tau)\,, 
\end{split}
\end{equation}
where in the second equality we apply the DFT on the two gravitational fields and use 
\begin{equation}
\sum_\mathbf{n} e^{ik_{\text{IR}} dx \mathbf{n} (\tilde{\mathbf{n}} - \tilde{\mathbf{n}}')} = N^3 \delta_{\tilde{\mathbf{n}} \tilde{\mathbf{n}}'}    .
\end{equation}

\begin{figure}[htb!]
    \centering
\includegraphics[width=0.95\linewidth]{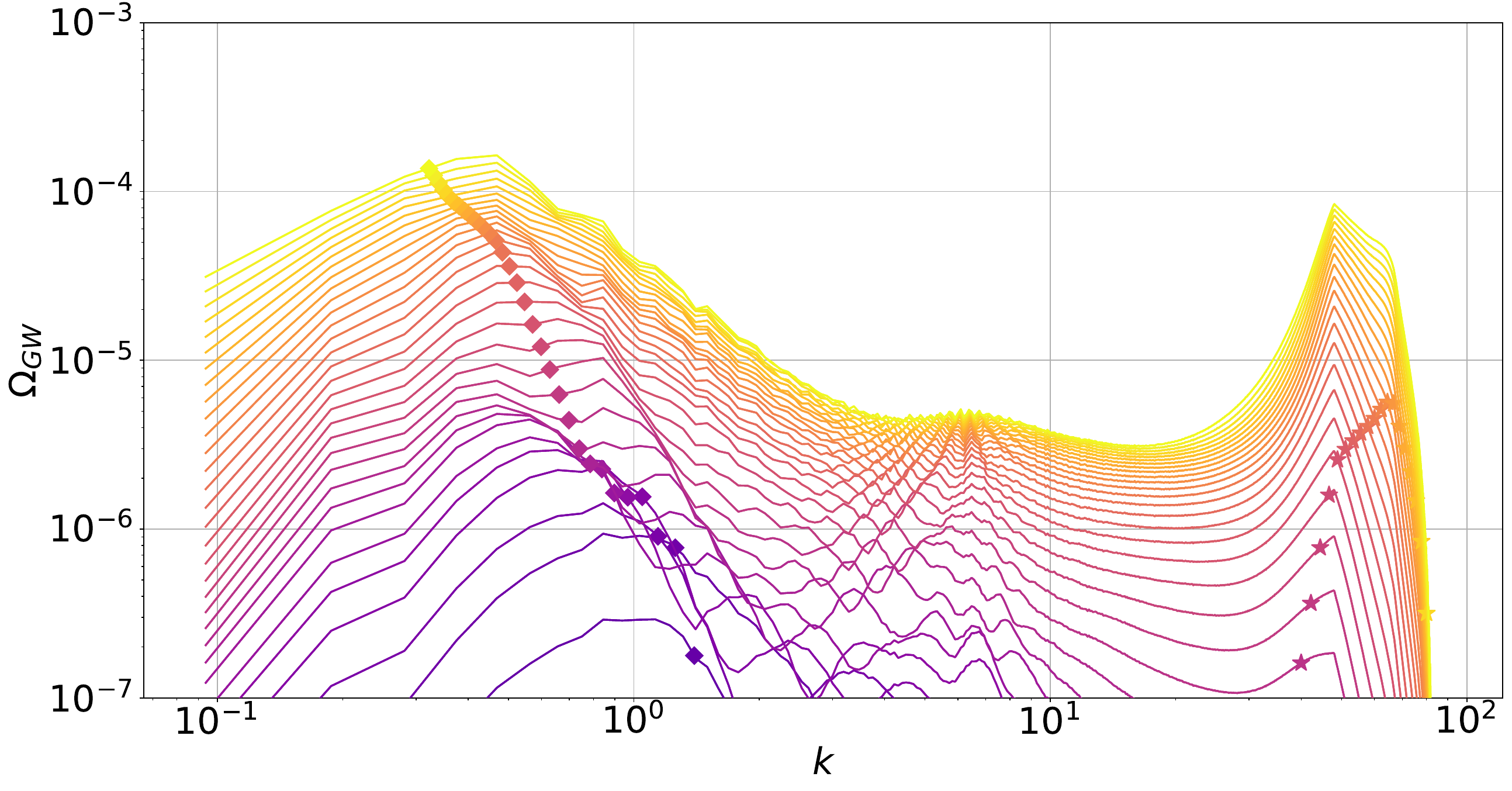}
    \caption{GW spectra in model with DW \eqref{matter-source} at subsequent conformal times from $\tau=1$ to $\tau=20$ with step \id{$\Delta\tau=1/2$}. Rhombs and stars indicate temporary conformal momenta corresponding to the Hubble parameter and the DW width, respectively.}
    \label{fig:Gw-with-wiggles}
\end{figure}

The problem is illustrated with plots in Fig.\,\ref{fig:Gw-with-wiggles}, 
where we present GW spectra, normalized to the total energy density $\rho_{tot}$ as 
\[
\Omega_{GW}\equiv \frac{1}{\rho_{tot}}\,\frac{d\rho_{GW}}{d \log k}\,,
\]
and calculated with code \CLns\, \cite{Figueroa:2021yhd,Figueroa:2023xmq} for the 
source given by the DW network formed by the scalar
field with the double well potential described by the lagrangian
\begin{equation}
  \label{matter-source}
{\cal L}=\half g^{\mu\nu} \d_\mu\phi \d_\nu \phi -\frac{\lambda}{4}\l
\phi^2-\eta^2\r^2.
\end{equation}
The conformal momentum $k$ is in units of $\sqrt{\lambda}\eta$ and conformal time $\tau$ is in units of $1/(\sqrt{\lambda}\eta)$. The running starts
at the conformal time $\tau=\tau_i=1$ when the scale factor equals   
$a=a_i=1$. Numerically we set $\eta=6\cdot 10^{16}$\,GeV, that allows for the rescaling of the results to any other values of the model parameters. The initial conditions for the
numerical simulations are chosen as $\phi(\tau_i)=0$ and the scalar
field fluctuations are \eqref{general-IC} with $\alpha=-3$, $k_{IR}=0$, $K_{UV}=1$ and  $T=T(\tau_i)/5$ where $T(\tau_i)$ is the initial temperature of the plasma.

In this case we exploit the so-called rigid (stable) domain walls, which after
formation quickly enter the scaling regime, when the same number of
domain walls are present inside the horizon patch. At each moment such a configuration 
produces GW with the peak frequency, determined by the Hubble momentum,
$k=2\pi a H(a)$ (marked on the spectra in Fig.\,\ref{fig:Gw-with-wiggles} by rhombs), and with a steadily growing amplitude. The worrisome  infrared part of
the spectrum close to the peak at a given time was  
mostly formed just a bit early, for details see e.g.\,\cite{Hiramatsu:2013qaa, Dankovsky:2024zvs}. As the simulation time grows, the Hubble shifts towards the lowest momenta, the infrared part with oscillating behavior shrinks, which at the late times hosts one oscillation only, that noticeably lifts up the infrared tail the spectra. 

The rigid walls must be totally destroyed, otherwise they overclose
the Universe\,\cite{Zeldovich:1974uw}. If the destruction is fast, say, it takes a Hubble time
or less, then the final spectrum looks as that on the plots of Fig.\,\ref{fig:Gw-with-wiggles} with largest $\tau$,  
and the infrared part exhibits the parasitic wiggles. The problem is
how to get rid of them and accurately infer the very part of the spectrum, which is heavily dependent on the dynamics of the DW destruction and therefore is actually of the primary interest in each realistic model.

{\bf 3.} The simplest way would be to proceed by running the code with
the matter source being switched off. Then sum up the infrared tail
of the spectrum calculated at several times and average. However, the
averaging procedure implies running the code over several periods, which for the oscillating
infrared tail exceeds tens of Hubble times at the moment of switching off
the source.

In the real lattice simulations the expanding Universe is described in the conformal (comoving) coordinates. Hence, while the running time  is limited from below by the time needed to enter a steady state regime (in the case of DW it is the scaling regime with fixed number of long DW inside the Hubble volume), it is also limited from above.  Indeed, if radiation or matter dominate in the Universe, the Hubble scale $l_H\propto 1/H$ grows faster than the scale factor $a$, which means that the Hubble volume in the comoving coordinates grows like $\l Ha\r^{-3}$ and finally exceeds the lattice volume $L^3$, which makes further simulations with generic periodical boundary conditions pointless. Another limiting process is associated with DW behavior: it's physical width $\delta_W=\sqrt{2/\lambda}/\eta$ is constant, and hence shrinks in the comoving coordinates as $a^{-1}$. Once its comoving size became smaller than the lattice spacing $L/N$, we stop to recognize DWs on the lattice, that again makes further simulations pointless.  

\begin{figure}[!htb]
    \centering
    \includegraphics[width=0.95\linewidth]{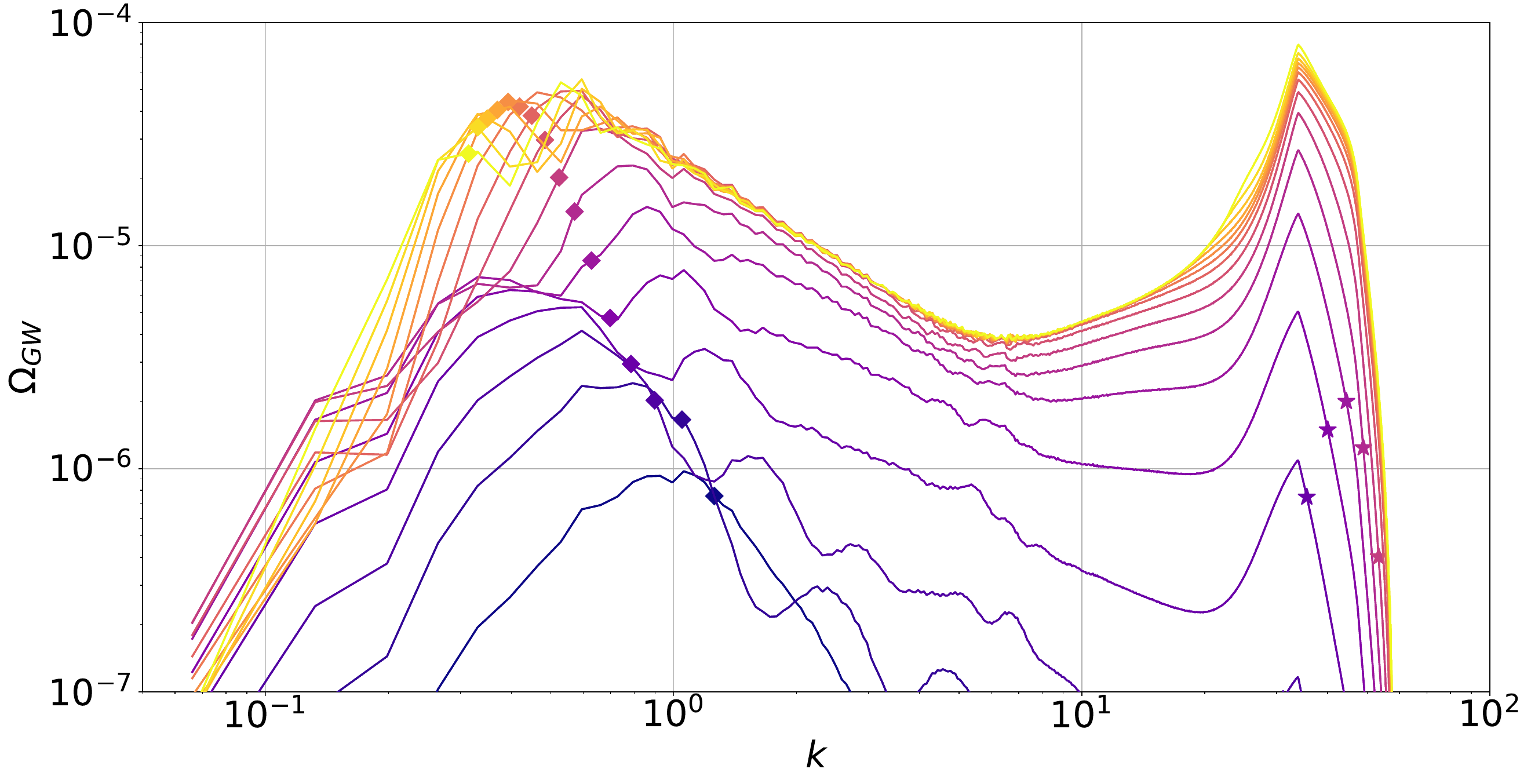}
    \caption{GW spectra at conformal times 
    from $\tau=5$ to $\tau=20$ with step $\Delta\tau=1$  (lines are indicated by colors changing from blue to yellow) in the model with DWs originated from the biased double-well potential, where DW network finally disappears; the GW spectrum is saturated by $\tau=20$. Rhombs and stars indicate temporary conformal momenta corresponding to the Hubble parameter and the DW width, respectively.}
    \label{fig:GW-with-bias}
\end{figure}

The problem with time limitation is much worse in case of models with realistic mechanisms destroying the DW network, since its proper operation also takes some time. The simplest case is when the scalar potential of eq.\,\eqref{matter-source} is supplemented with a bias, that is a small term explicitly breaking $Z_2$-symmetry, e.g. $\epsilon \phi^3$, $\epsilon\ll \eta$. This term removes the degeneracy between the two vacuums, so the DW network dissolves when the true vacuum occupies the whole space. The example of spectrum of GW produced by such a dissolving DW with $\epsilon=0.025\lambda\eta$ is presented in Fig.\,\ref{fig:GW-with-bias}. 
See also Refs.~\cite{Kitajima:2023cek, Cyr:2025nzf, Notari:2025kqq, Babichev:2025stm, Barbini:2026edx} for earlier simulations of GWs from biased DWs. 
One observes that wiggles in the infrared part of the spectrum is even more pronounced, than those in case of rigid DW in Fig.\,\ref{fig:Gw-with-wiggles}. 

\begin{figure}[!htb]
    \centering
    \includegraphics[width=0.95\linewidth]{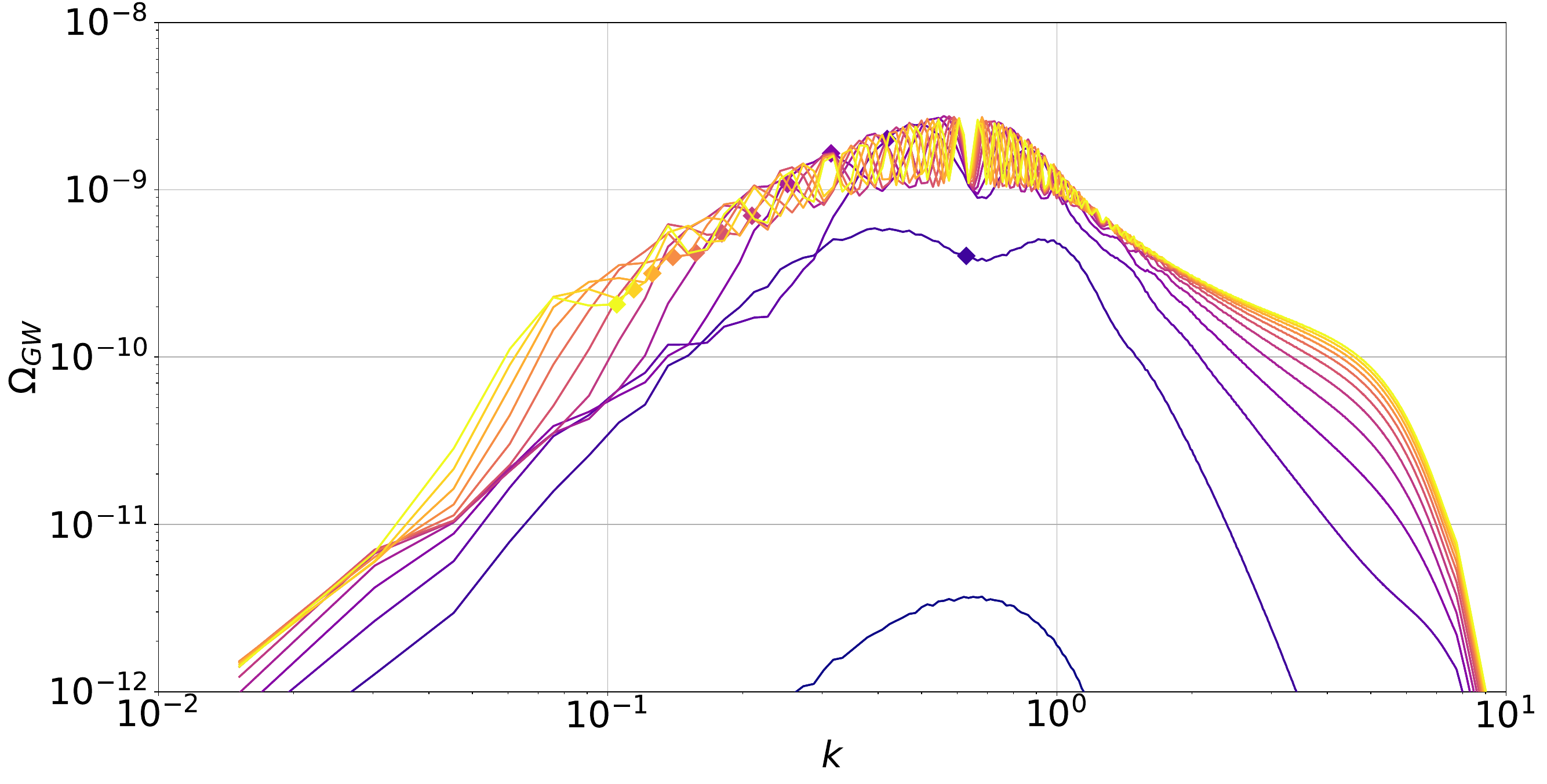}
    \caption{Typical prediction of GW spectra produced by melting domain walls \cite{Ramazanov:2021eya,Babichev:2021uvl,Dankovsky:2024ipq}; conformal times run from 5 to 60 with step $\Delta\tau=5$, the color of corresponding lines varies from blue to yellow.}
    \label{fig:melting}
\end{figure}

Another mechanism of destruction of DW is provided by models exhibiting an inverse phase transition, when $Z_2$ symmetry gets restored in the late Universe. It can be realized, e.g., with mass term added to the scalar potential \eqref{matter-source} and parameter $\eta$ proportional to the plasma temperature, which decreases with the Universe expansion. Then the potential barrier and the distance between the two vacuums decrease, and DWs melt in time \cite{Ramazanov:2021eya,Babichev:2021uvl}. In this case the GW are extensively produced right after $Z_2$-breaking, and the peak frequency of GW are determined by the Hubble parameter of that epoch \cite{Dankovsky:2024ipq}. The typical spectrum predicted in such models is presented in Fig.\,\ref{fig:melting}.

\begin{figure}
    \centering
    \includegraphics[width=0.95\linewidth]{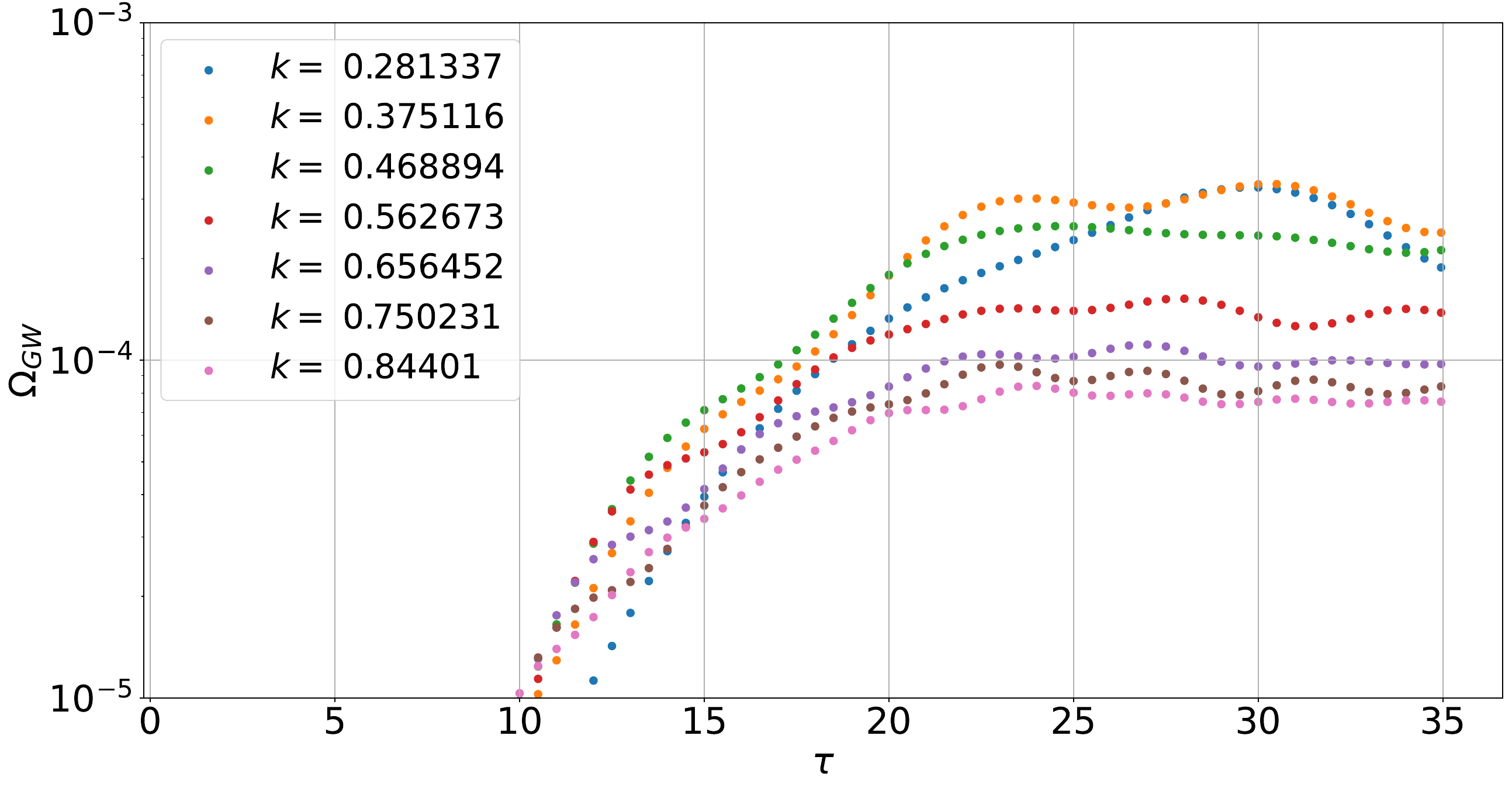}

    \includegraphics[width=0.95\linewidth]{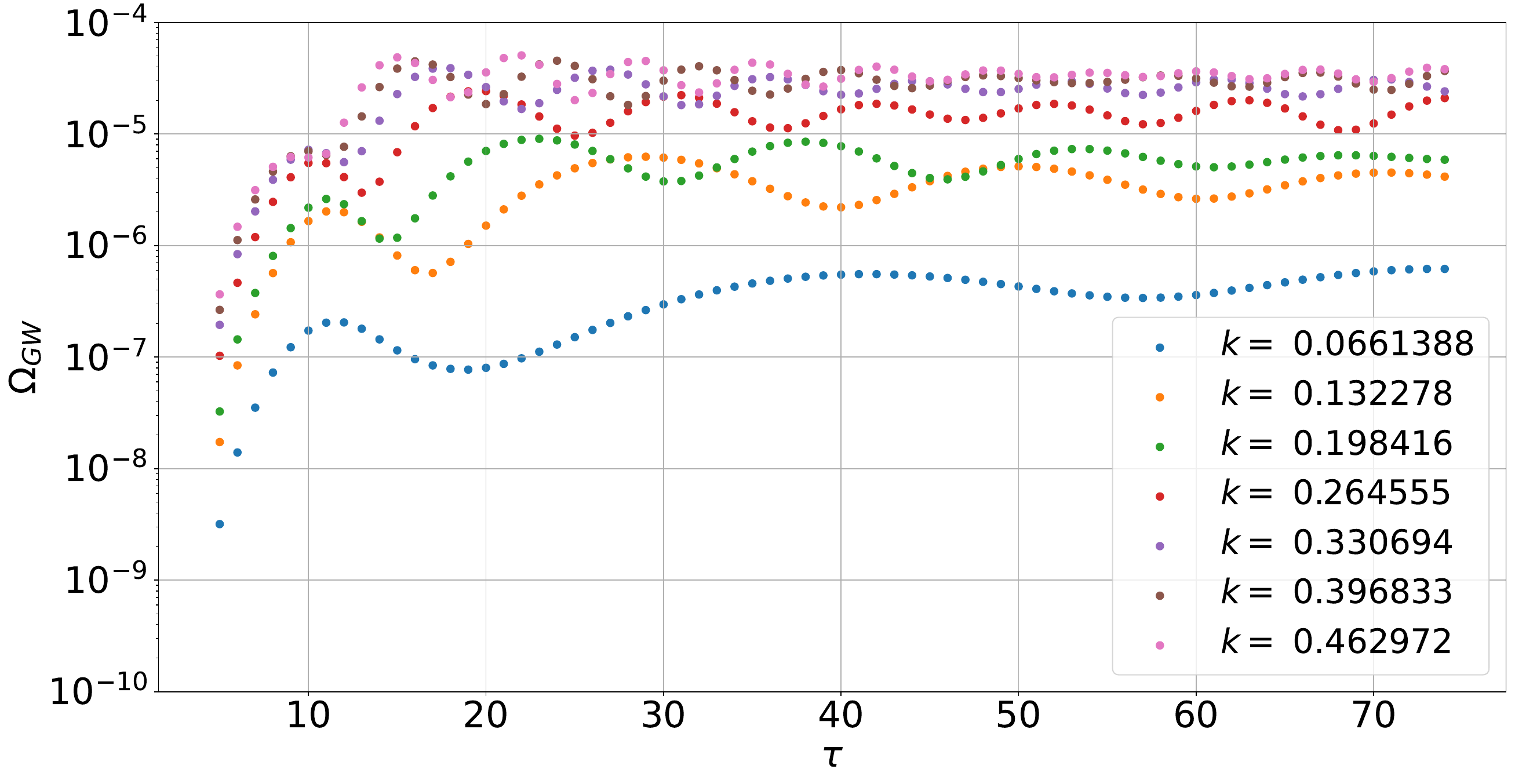}
    \caption{Time evolution of GW amplitudes for a set of modes. {\it Top panel:} the case of rigid domain walls. {\it Bottom panel:} the case of biased potential with $\epsilon=0.025\lambda\eta$. }
    \label{fig:mode-oscillations}
\end{figure}

{\bf 4.} To deal with the wiggly tails of GW spectra we suggest to consider the interesting infrared modes and average separately the amplitude of each mode over some time interval right before finishing the simulations. Indeed, in each numerical simulation we deal with a fixed set of conformal momenta, and the amplitude of each mode from the infrared tail exhibits oscillations, see Fig.\ref{fig:mode-oscillations}. 
In case of biased scalar potential, where the DWs dissolve, so the GW source disappears, the late time asymptotics of each amplitude are transparent. Even in the case of rigid domain walls, when the source operates till the end of simulations, the expected asymptotics of most infrared modes can be inferred upon averaging (and for the lowest mode may be with some extrapolation). 

\begin{figure}
    \centering
    \includegraphics[width=0.95\linewidth]{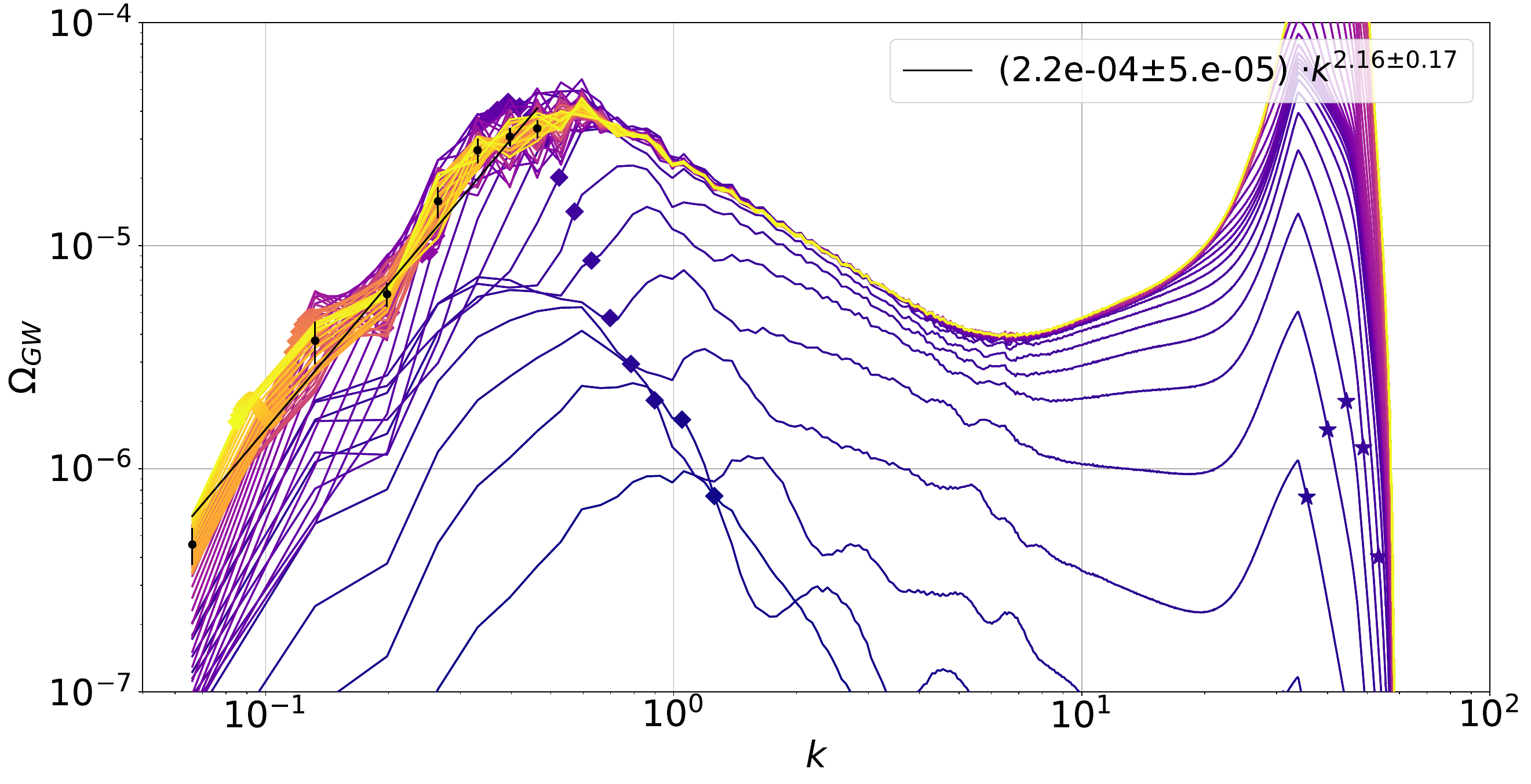}
    \caption{
    GW spectra for the case of biased potential with $\epsilon=0.025\lambda\eta$, where the GW production stops at $\tau=20$. The same model as in Fig.\,\ref{fig:GW-with-bias}, but the simulations proceeds to $\tau=75$ and the spectra are presented for the times separated by $\Delta\tau=1$.  The wiggling IR modes, presented on bottom panel of Ref.\,\ref{fig:mode-oscillations}, are separately averaged over time as described by eq.\,\eqref{averaging}. Thus averaged IR part is fitted with a simple power law. \id{$\chi^2_{\nu} = 3.65$, $N=5$}}
    \label{fig:GW-averaged-spectrum}
\end{figure}

We suggest estimating the asymptotic amplitudes with simple averaging given by the following formula
\begin{equation}
\label{averaging}
    \Omega_{GW,avg} \left( k \right)   = \frac{1}{N_k} \sum_{i=1}^{N_k} \Omega \left( k, \tau_i \right) \, ,
\end{equation}
where for each mode $k$ the set $\{\tau_i\}$ contains  subsequent $N_k$ time points (separated by the time step of simulations) between the latest two local peaks of the amplitude $\Omega_{GW} \left( k, \tau \right)$, see Fig.\ref{fig:mode-oscillations}. This simple procedure gives a good result, illustrated in Fig.\,\ref{fig:GW-averaged-spectrum}, 
where we present results for the scalar model with biased potential, where the DW network evolves for a limited time
before destruction, and the simulations proceed for a few Hubble time intervals after. The error bars are estimated by calculating the standard deviation of $\Omega \left( k, \tau_i \right)$ from its average value $\Omega_{GW,avg} \left( k \right)$ for the set 
of time points $\{\tau_i\}$. The fitting to the infrared tail of the GW spectrum is performed by making use of the function curve\_fit from the Python library scipy.

\begin{figure}
    \centering
    \includegraphics[width=0.95\linewidth]{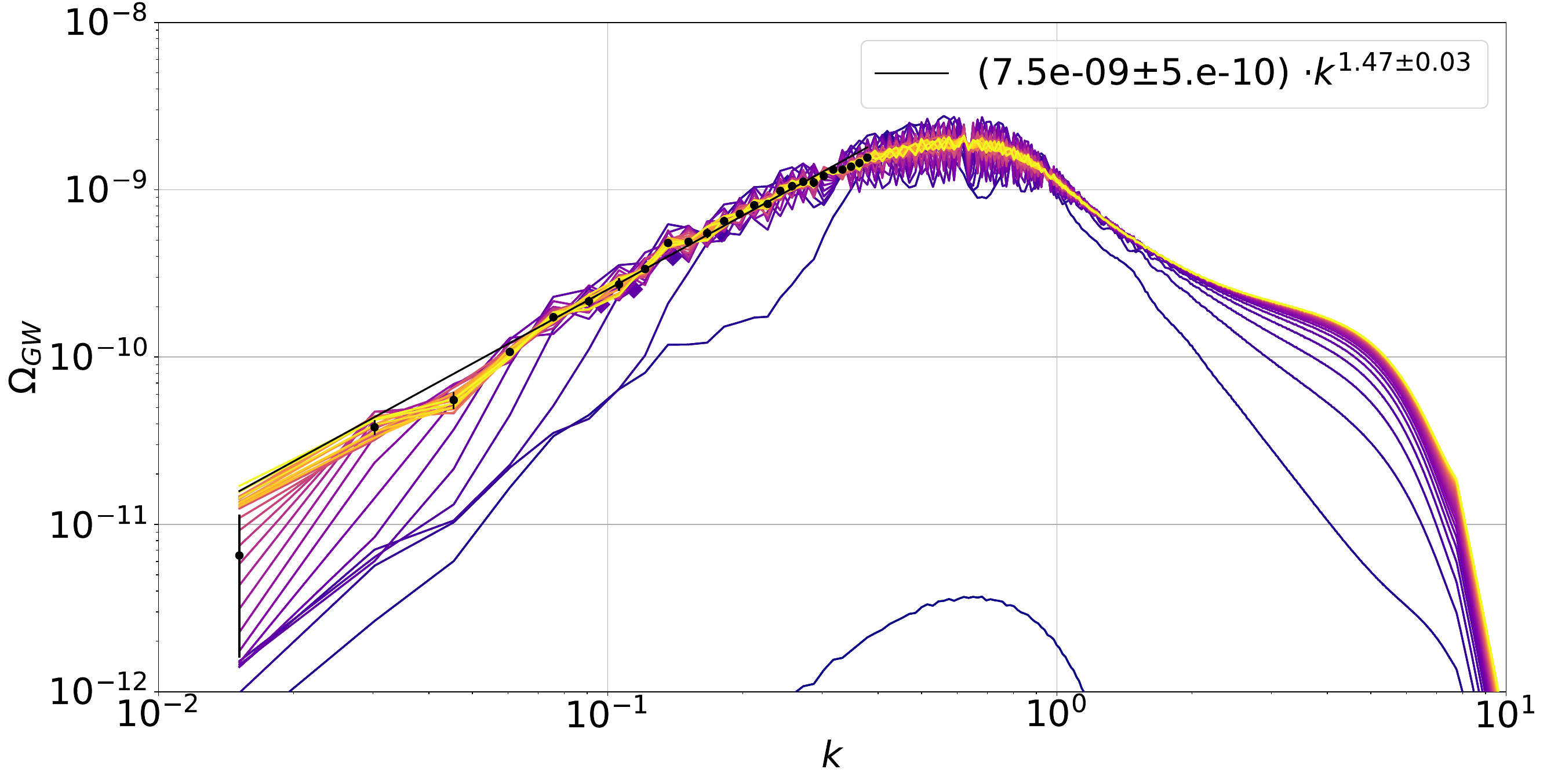}
    \caption{
    GW spectra for the case of melting domain walls, where the GW production terminates at $\tau=60$. The same as Fig.\,\ref{fig:melting}, but with extended simulations. The conformal time runs from $\tau=5$ to 300, and the GW spectra are presented with the step $\Delta\tau=10$.  The wiggling IR modes are separately averaged over time as described by eq.\,\eqref{averaging}. Thus averaged IR part is fitted with a simple power law. 
    \id{$\chi^2_{\nu} = 3.77$, $N=23$}
    }
    \label{fig:GW-averaged-melting}
\end{figure}

Results of similar averaging of IR modes in case of melting DWs, see Fig.\,\ref{fig:melting}, and the subsequent fit with simple power law to the IR part of the GW spectrum, are presented in Fig.\,\ref{fig:GW-averaged-melting}. 

\begin{figure}
    \centering
    \includegraphics[width=0.95\linewidth]{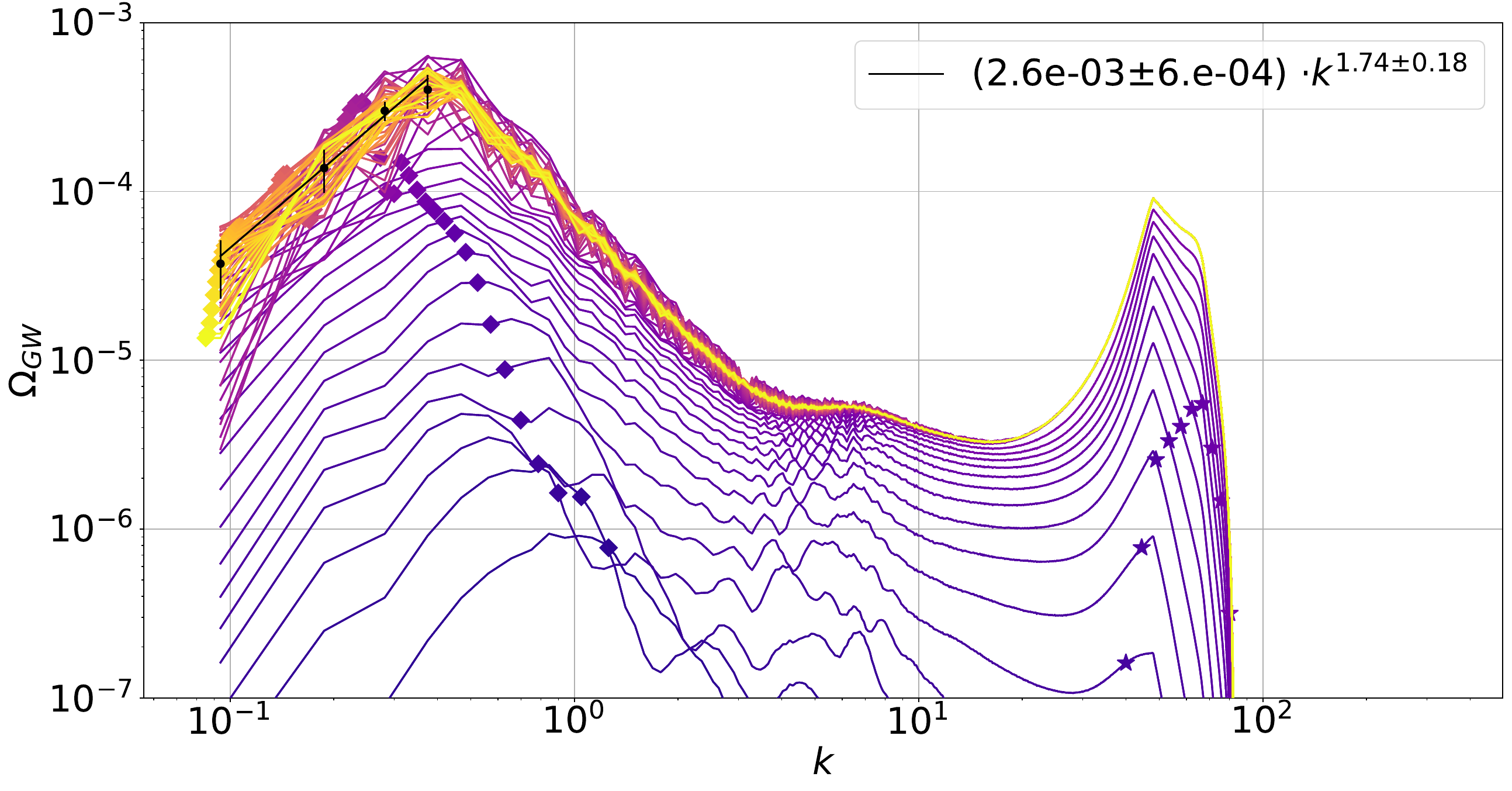}
    \caption{
    GW spectra for the case of rigid domain walls, 
    assuming their rapid destruction at $\tau=20$. The same as Fig.\,\ref{fig:Gw-with-wiggles}, but with extended simulations. The conformal time runs from $\tau=1$ to 75, and the GW spectra are presented with the step $\Delta\tau=1$.  The wiggling IR modes are separately averaged over time as described by eq.\,\eqref{averaging}. Thus averaged IR part is fitted with a simple power law. 
    \id{$\chi^2_{\nu} = 0.38$, $N=2$}
    }
    \label{fig:GW-averaged-rigid}
\end{figure}

The case of rigid domain walls is a toy model, since in any realistic case the DW network must be finally destroyed. The final spectrum of GW is saturated mostly at the moment of DW destruction and hence  generically depends on the destruction mechanism. To illustrate the applicability of our averaging procedure in this case, let us assume that the destruction happens simultaneously, e.g. takes much less than a Hubble time. Then we can mimic this in the numerical simulations by switching off the scalar source at a given time with a step-function   
and proceeding with simulations for some time. Then we stop simulations and perform the averaging that yields the spectrum presented in Fig.\,\ref{fig:GW-averaged-rigid}.  

{\bf 5.} 
The described above procedure requires additional operation time, which may be quite long (not a few, but several Hubble times) to achieve proper averaging of the modes of the lowest momenta, especially if they are produced at the latest moments of simulations.   
A potentially promising way to increase the running time of simulations by modifying the scalar field equation was presented in Ref.\,\cite{Press:1989yh}. The equation of motion for the scalar field $\phi$ with potential $V(\phi)$ in the expanding Universe can be written as 
\[
\phi''+\gamma\,\frac{a'}{a}\,\phi'-\Delta\phi +a^\beta\,\frac{dV(\phi)}{d\phi}=0\,,
\]
with dimensionless parameters fixed as 
$$
\gamma=\beta=2\,.
$$ 
This equation follows directly from the lagrangian\,\eqref{matter-source}. Alternatively, Ref.\,\cite{Press:1989yh} suggested another choice,  for the radiation dominated stage it reads 
\begin{equation}
\label{ALT}
    \gamma=3\,,\;\;\; \beta=0\,.
\end{equation}    
It makes the DW width constant in the comoving coordinates, yet keeps the DW tension constant and non-relativistic matter density $\rho_m$ (which the scalar particles form in the bulk, i.e. outside the DW) decreasing with the scale factor as in the realistic case, i.e. $\rho_m\propto a^{-3}$. In such a case the DW width places no upper limits on the simulation time, while the cosmological behavior of the massive scalar particles in the bulk (out of the DWs) and DW dynamics in the limit of infinitely thin DW (the Nambu-Goto regime) are correctly reproduced. Then one can expect some artificial dynamics of the scalar ultra-violet modes on the lattice, but generally correct behavior of the infra-red modes. And longer numerical simulations would allow for taming the wiggles in the infrared part of GW spectrum. This approach was investigated in e.g. Refs.\,\cite{Larsson:1996sp,Sousa:2010zza} and is used 
occasionally in later numerical studies. 

\begin{figure}[!htb]
    \centering
    \includegraphics[width=0.95\linewidth]{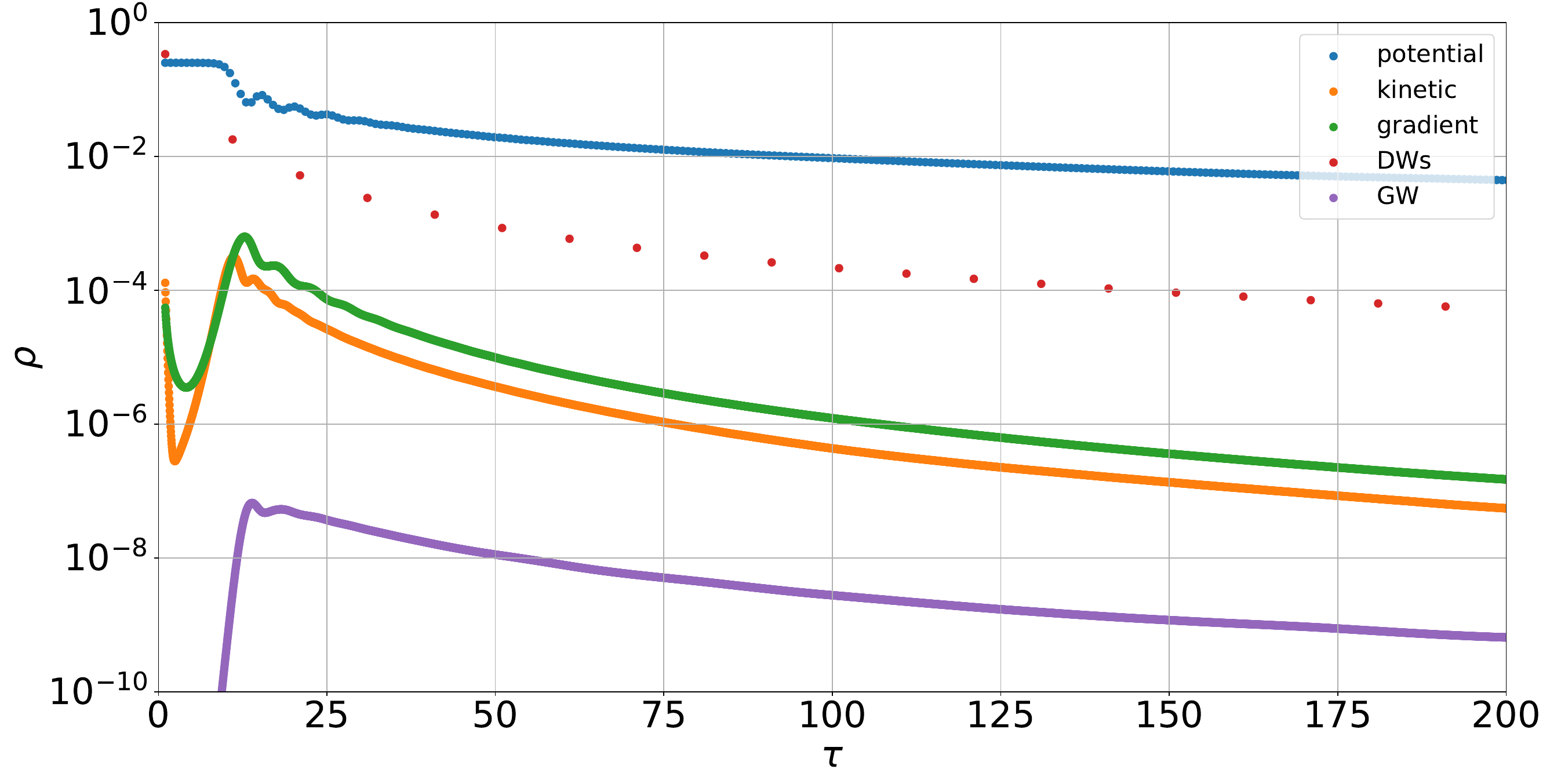}
    \caption{Time evolution of GW energy density and various terms of scalar field energy density in the expanding Universe in the model with the scalar field equation rescaled as \eqref{ALT}.} 
    \label{fig:PRS-failed}
\end{figure}

However, we find the scaling \eqref{ALT}  inappropriate for numerical calculations of the GWs produced by the DW network: the obtained spectrum is wrong both in the ultra-violet and infrared parts. Not spectra, but even the behavior of the GW energy density with the scale factor is wrong, as we illustrate in Fig.\,\ref{fig:PRS-failed} 
for the case of rigid walls and the same initial scalar perturbations as in Fig.\,\ref{fig:Gw-with-wiggles}. 
For the scaling solution of the DW network (constant number of long DWs inside the Hubble volume) its energy density scales as $\rho\propto a^{-2}$. Then eq.\,\eqref{GW-equation} implies that the gravitational amplitude grows quadratically with the scale factor, and hence the GW energy density \eqref{GW-energy} remains constant. The energy density of GW in Fig.\eqref{fig:PRS-failed}, obtained with scaling \eqref{ALT} decreases in time.    

This result is not quite unexpected, since the scaling\,\eqref{ALT} changes the evolution of gradient term in the scalar energy density, which sources the GW\,\eqref{GW-equation}. The evolution of the different components of the scalar density is also presented in Fig.\,\ref{fig:PRS-failed}. 
One observes that while the DW energy density evolves correctly with  the scale factor, the total scalar energy density is determined by the scalar potential: it dominates over the DW contribution, the kinetic and gradient terms, which both decrease as $\propto a^{-3}$.

\begin{figure}[!htb]
    \centering
\includegraphics[width=0.95\linewidth]{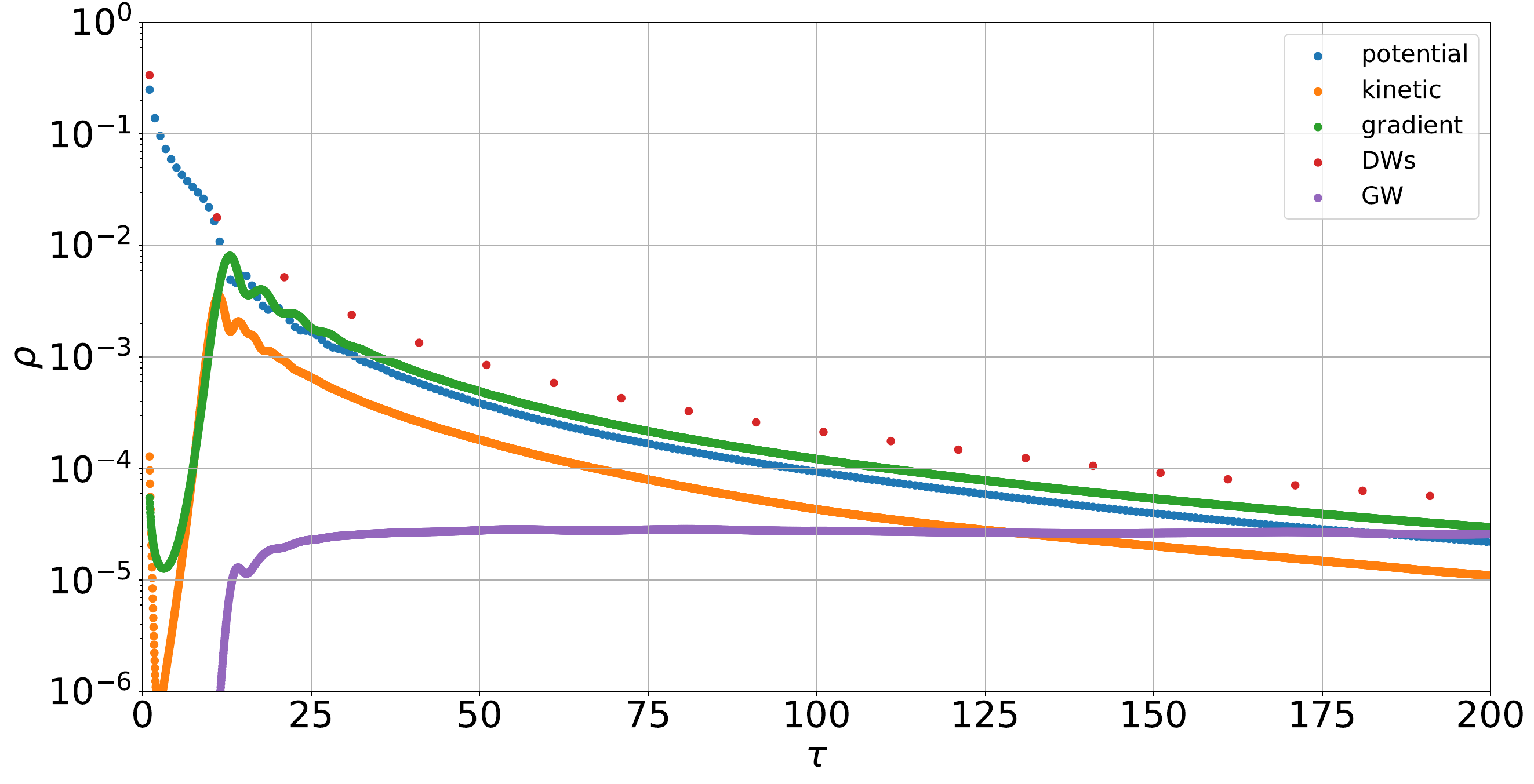}
    \caption{The same energy density components as in Fig.\,\ref{fig:PRS-failed} but rescaled with powers of the scale factor $a(\tau)$ to mimic their correct behavior in the physical case without rescaling \eqref{ALT}.}
    \label{fig:rescaling}
\end{figure}

\begin{figure}[!htb]
    \centering
\includegraphics[width=0.95\linewidth]{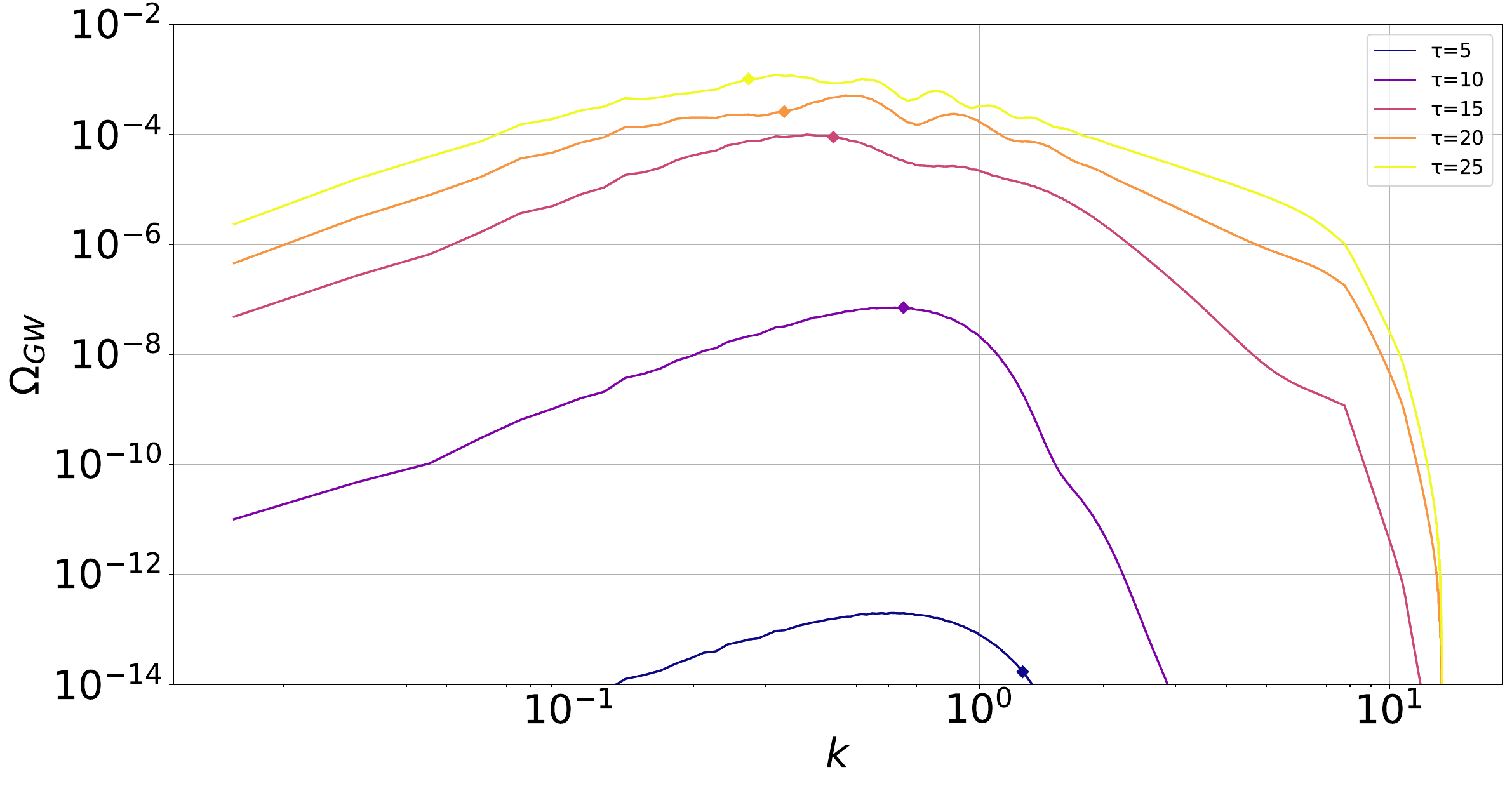}

   \includegraphics[width=0.95\linewidth]{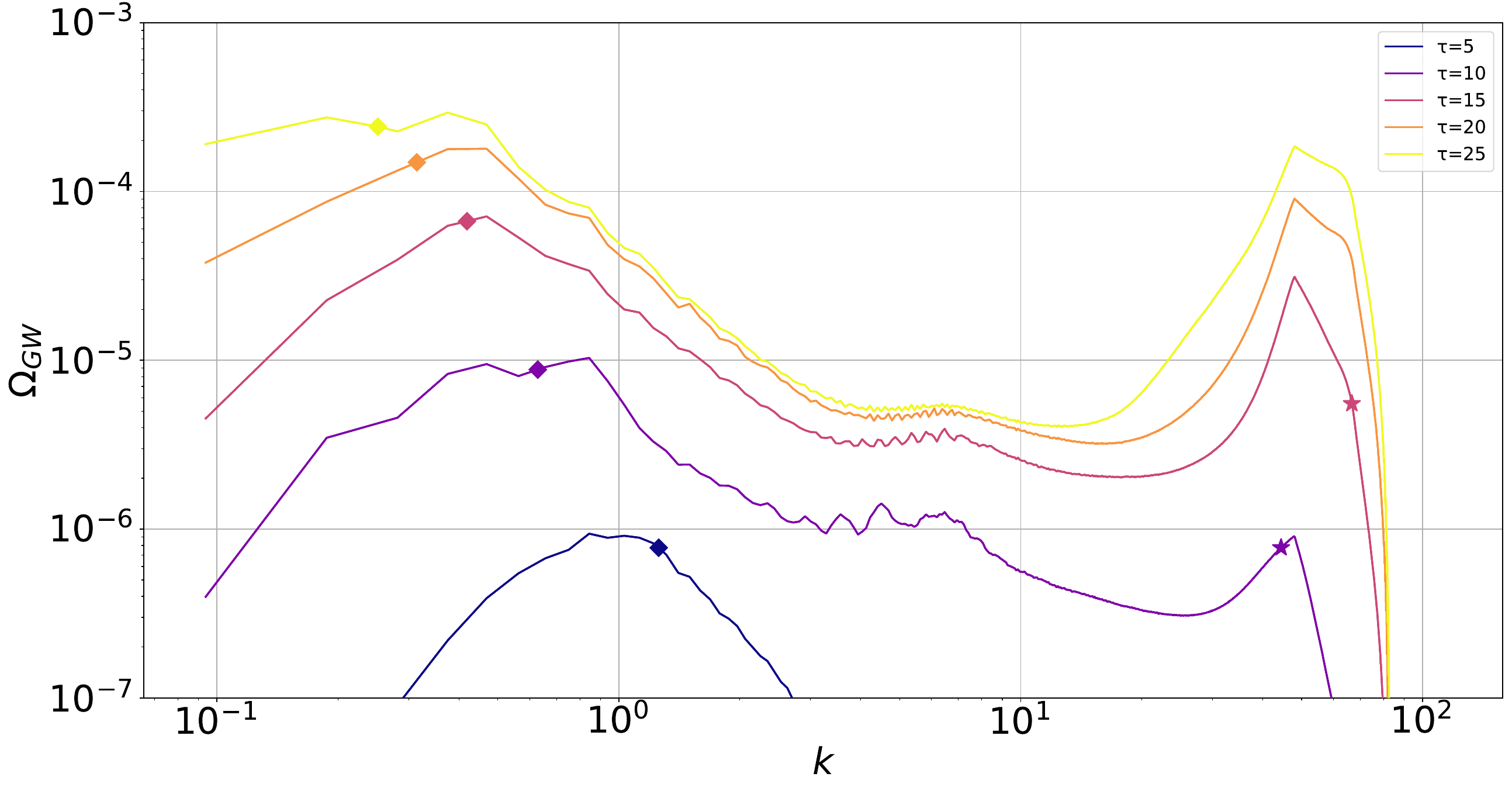}
    \caption{GW spectra calculated with the same physical parameters of the scalar potential and initial perturbations but with rescaling \eqref{ALT} and further corrected energy densities as in Fig.\,\ref{fig:rescaling} ({\it top panel}) and without any rescaling and corrections ({\it bottom panel}).}
    \label{fig:spectra-with-rescaling}
\end{figure}

One may try "to correct" for this wrong scaling by artificially multiplying the scalar energy density terms by properly chosen powers of the scale factor, restoring their correct behavior with the scale factor, see Fig.\,\ref{fig:rescaling}.  
The GW density then remains constant in time, as it must, see Fig.\,\ref{fig:rescaling}. However, the GW spectra, see the top plot of Fig.\,\ref{fig:spectra-with-rescaling}, 
are still not what we obtained without the scaling\,\eqref{ALT}, see the bottom plot of Fig.\,\ref{fig:spectra-with-rescaling}.

{\bf 6.} To summarize, we consider the problem of getting rid of the parasitic wiggles observed in the numerical calculations of the GW spectra at low frequencies. The problem is illustrated on example of the DW network producing the GW in the early Universe. We show that the artificial scaling of different parts in 
the scalar equation yields evidently wrong spectra. Alternatively, we suggest to average the amplitude of each mode over some latest time interval of numerical simulations. The obtained values and the corresponding dispersions can be straightforwardly fitted with simple power law, which gives the robust predictions to be tested at experiments on searches and measurements of GWs. 

\vskip 0.3cm 
{\it Acknowledgments.} We thank S.\,Ramazanov for a valuable discussion. Numerical simulations have been executed on the cluster of the Theoretical Division of INR RAS. The work of I.~D. and D.~G. was supported in part by the scientific program of the National Center for Physics and Mathematics, section 5 "Particle Physics and Cosmology", stage 2026-2027. I. D. acknowledges the support by the Foundation for the Advancement of Theoretical Physics and Mathematics "BASIS".

\vskip 1cm

\bibliographystyle{utphys}
\bibliography{refs}
\end{document}